\documentclass[3p,12pt]{elsarticle}

\usepackage{amssymb}
\journal{Materials Engineer and Science A}

\begin{document}

\begin{frontmatter}

\title{Deformation of Palladium Plates by a Small External Stress during Hydrogen Absorption and Desorption}

\author[label1]{Akio Kawasaki\fnref{CurAdd}}
\ead{akiok@mit.edu}
\fntext[CurAdd]{Present Address: Department of Physics, Massachusetts Institute of Technology, Cambridge, MA 02139, USA}
\author[label1,label2]{Satoshi Itoh}
\author[label3]{Kunihiro Shima}
\author[label1,label2]{Toshimitsu Yamazaki}
\address[label1]{Department of Physics, University of Tokyo, Bunkyo, Tokyo 113-0033, Japan}
\address[label2]{RIKEN, Nishina Center, Wako, Saitama-ken 351-0198, Japan}
\address[label3]{Tanaka Kikinzoku Kogyo K.K., Tomioka, Gunma-ken 370-2452, Japan}

\begin{abstract}
We observed an unexpected bending of a horizontally held palladium plate at 150$^{\circ}$C under a small external shearing stress. The deformation was fastest at the nominal composition of PdH$_{{\rm 0.1-0.2}}$, which is in the $\alpha+\beta$ phase.  We also observed that a plate warped back and forth during a cycle of absorption and desoprtion of hydrogen when it was hung vertically, which does not have a reasonable explanation.

\end{abstract}


\end{frontmatter}


\section{Introduction}
\begin{figure}[!b]
 \begin{center}
 \includegraphics[width=7.5cm ,clip]{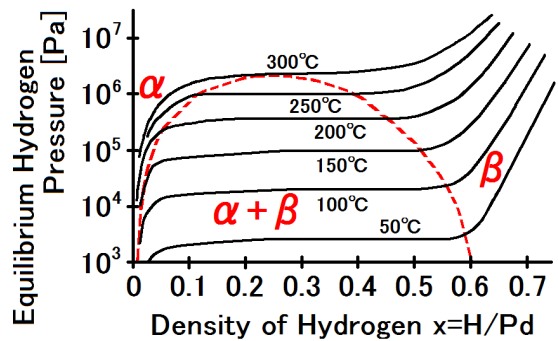}
  \caption{The phase diagram of the Pd/H system according to Ref. \cite{Phase,HinM} : the $\alpha$, $\beta$ and their mixed phases are indicated together with a phase boundary by a broken curve.} 
  \label{phaseDiagram}
 \end{center}
\end{figure}

Palladium metal is well known to absorb an abundant amount of hydrogen, and is widely used for filtering and storage of hydrogen \cite{Pd-H}.  When palladium metal is saturated with hydrogen, the ratio of the hydrogen atom to the palladium atom is about PdH$_{0.6}$, depending on the temperature and the pressure of the hydrogen gas \cite{Phase,HinM}, as shown in Fig.\ref{phaseDiagram}.

Pure palladium metal with an fcc lattice changes into the $\alpha$ and $\beta$ phases, and also into  their mixed phase, $\alpha + \beta$, when it absorbs hydrogen atoms.  Hydrogen atoms are in the ionized form and are located at octahedral interstitial sites in both phases \cite{alpha,beta}. The two phases differ in the concentration of hydrogen atoms to palladium atoms, corresponding to the the difference in the lattice constant: 3.89\AA \ for the $\alpha$ phase and 4.02\AA \ for the $\beta$ phase \cite{Sugeno}.  This corresponds to the volume increase when it absorbs hydrogen and to the volume decrease during the desorption, but the bulk does not return to its original shape after desorption.  Many irreversible deformations have been reported \cite{OldPaper,PMR,ST,IJHE, AM,JSMS1,JSMS2}.  

Recently, palladium metal of various shapes was found to shrink spontaneously in the direction of minimizing the surface area after several cycles of absorption and desorption of hydrogen at 150$^{\circ}$C \cite{JpnAcad}. The observed drastic changes of Pd shape seemed to indicate that an enormous stress was produced to cause such a deformation, if the metal kept the ordinary elasticity.  Another explanation to this phenomenon is superplasticity.  Superplasticity in general has been studied for more than 50 years and it is characterized as large plastic deformation of several hundreds percent strain happening in alloys with small crystal grain.  For palladium, it is reported in Ref.\cite{Gortsov1981} for the first time.  

Beeri et al. investigated the change in the strain of a palladium wire during the absorption and desorption of hydrogen in terms of superplasticity. \cite{Beeri2009}  Their experiment was conducted under room temperature and showed the result consistent with Ref.\cite{JpnAcad} in the sense that they observed elongation during the absorption and shrinkage after one cycle of absorption and desorption of a wire when the stress on it was relatively small.  It was, however, designed to study the deformation of the wire only in one dimension, and no information was obtained as to at which concentration of hydrogen such deformation occurs. 

\begin{figure}[!tb]
 \begin{center}
  \includegraphics[width=7.5cm,clip]{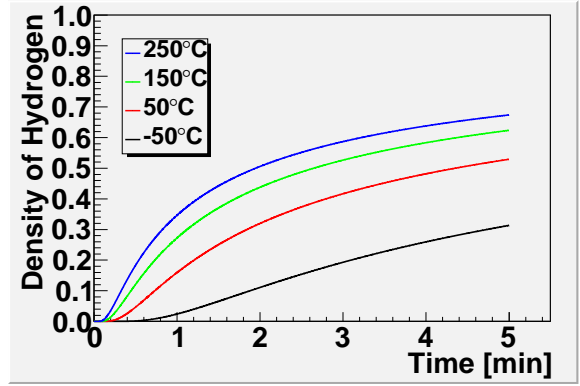}
  \caption{Density of hydrogen atoms at the center of the 1 mm thick plate: this was calculated by solving a one-dimensional diffusion equation with the initial condition that the density of hydrogen is 0 inside the plate and 1 outside the plate.}
  \label{diffusion}
 \end{center}
\end{figure}

In our experiment, we held a Pd plate horizontally and performed a cycle of hydrogen absorption and desorption.  This setting allowed the plate to undergo one dimensional deformation, bending upward or downward, and was sensitive to the deformation even if the stress was small.  We aimed at obtaining in-situ information on the dynamical change of the shape of a Pd plate during its absorption and desorption of H$_2$ gas by measuring the transformation of a plate of palladium with weights hung to it.  We set the temperature at 150$^{\circ}$C and took snap-shot photographs.  The temperature 150$^{\circ}$C was high enough for a sufficient amount of hydrogen to diffuse to the center of the plate in a short time, as shown in Fig. \ref{diffusion}.  A calculation from the permeability constant shows that this speed is not limited by the transportation on the Pd surface.  If we assume this absorption as a permeation of hydrogen to 0.2 mm thick Pd bulk through 0.4 mm Pd plate, a calculation from permeability constant in \cite{Morreale2003} shows that the maximum increase rate in the hydrogen amount is $7.6 \times 10^{-5}$ mol/s, whereas it is $2.3 \times 10^{-7}$ mol/s in Fig. \ref{diffusion}.  Also, our setting enabled us to relate the amount of the hydrogen absorbed to the deformation.  In addition to the measurement with horizontally held Pd plate, we performed experiments with a vertically hung plate to get some other information on the behavior of the plate.

\section{Dynamical Measurements of the Bending of Pd Plates}
\subsection{Experimental Procedure: Bend}

\begin{figure}[!tb]
 \begin{center}
  \includegraphics[width=7.5cm,clip]{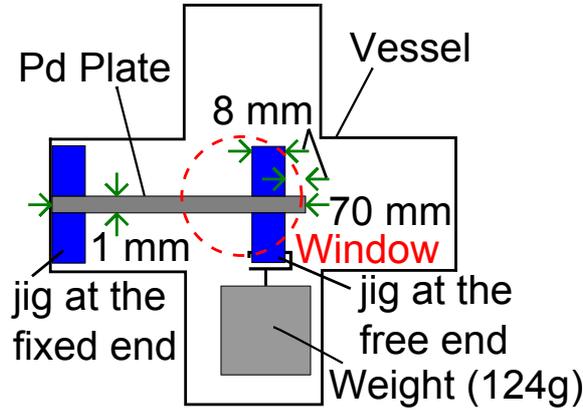}
  \caption{The setting for the experiments to measure the bending }
  \label{Horizontal}
 \end{center}
\end{figure}

A Pd plate having a rectangular shape (1 mm thick, 10 mm wide and 70 mm long) was held horizontally in a vacuum vessel, with one end of the strip fixed by a jig, and with a weight at the other end, hung as shown in Fig.~\ref{Horizontal}. The Pd plate had been annealed at 1000$^{\circ}$C for an hour before the experiments. The total weight of a jig and a gold weight was 124 g, which applied 122 kPa shearing force on the plate. The jig was fixed at 8 mm distance from the free end of the plate. 

The movement of the free end of the Pd plate was observed through a transparent window, as shown in Fig.~\ref{Horizontal}, and was recorded by a digital camera continuously during a hydrogen absorption/desorption cycle. A scaling pad with 1-mm thin mesh and 5-mm bold mesh was located 7.5 mm away from the plate to precisely measure the movement. The distance between the plate and the lens of the camera was about 30 cm. The precision for a measured displacement of the open end was as good as 0.1 mm.

The inner volume of the vessel that was filled with gas was 392 cm$^3$.  
The vessel was heated and kept at 150$^{\circ }$C. The vessel and the heaters, except for the transparent window, was wrapped with aluminum foil to maintain the temperature.

For each run, the plate was subjected to one cycle of absorption and desorption of hydrogen in the  following way.  First, we evacuated the vessel with a rotary pump and a turbo pump. Just after the evacuation started, we turned on heaters, one of which was controlled by a feedback system, so that the vessel would be kept at 150 $^{\circ}$C.  After the vacuum reached to the order of $10^{-2}$ Pa, we introduced hydrogen gas up to 1 MPa or 0.2 MPa.  When we set the pressure 1 MPa, we introduced the gas at once to 1.2-1.4 MPa, and waited until the pressure reached a constant value. When we introduced hydrogen up to 0.2 MPa, we gradually filled the vessel with hydrogen gas with a flow controller for D$_2$. We repeatedly added the gas once in 30 minutes or so until it reached a pressure of around 0.2 MPa. 

After the saturation was reached for a given gas pressure, we purged the gas to atmosphere, and then evacuated the vessel, first by a rotary pump, and then by a turbo pump.  In order for hydrogen to desorb from the palladium metal, we waited until the vacuum level reached the order of $10^{-4}$ Pa. This ensured that most of hydrogen was properly taken away, because the decrease of the pressure once became faster during evacuation to this pressure.  After each cycle, we took out the metal to examine its outlook and to adjust it for the next experiment.  

During these processes, the vacuum level of the system, the temperature of the vessel and the pressure of the vessel were recorded every 10 seconds by a LabVIEW system.  The state of the movement was recorded by taking photographs.  The interval of photo taking depended on the speed of deformation of the plate.

\subsection{Experimental Results: Bend}
\begin{figure}[!tb]
 \begin{center}
  \includegraphics[width=8.5cm,clip]{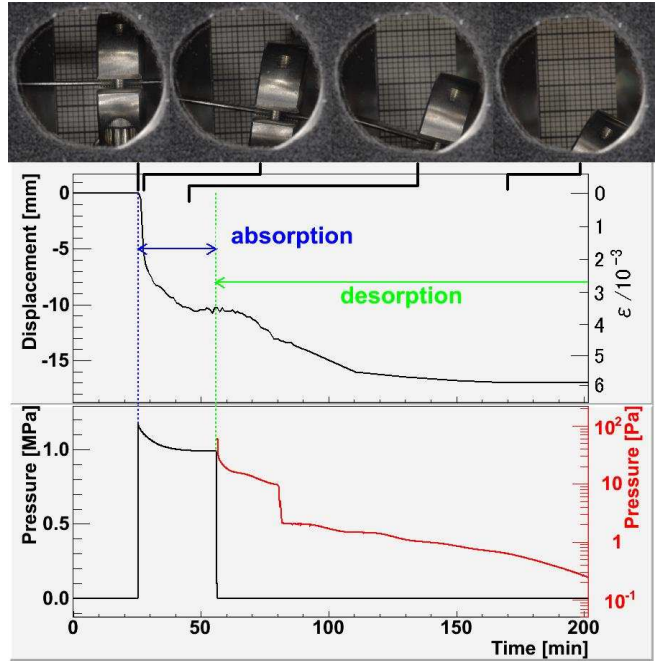}
  \caption{Result B1; 1 MPa H$_2$}
  \label{Pressure0218}
 \end{center}
\end{figure}

\begin{figure}[!tb]
 \begin{center}
  \includegraphics[width=8.5cm,clip]{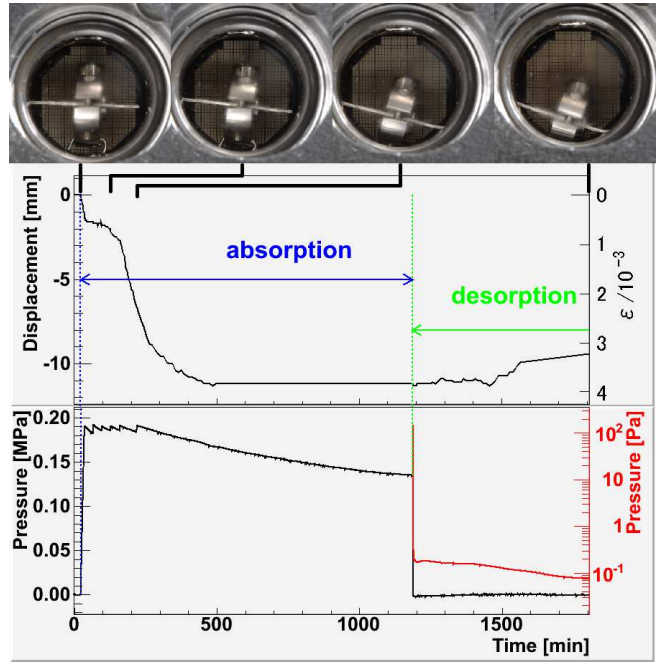}
  \caption{Result B2;  0.2 MPa H$_2$}
  \label{Pressure0318}
 \end{center}
\end{figure}

\begin{figure}[!tb]
 \begin{center}
  \includegraphics[width=7.5cm,clip]{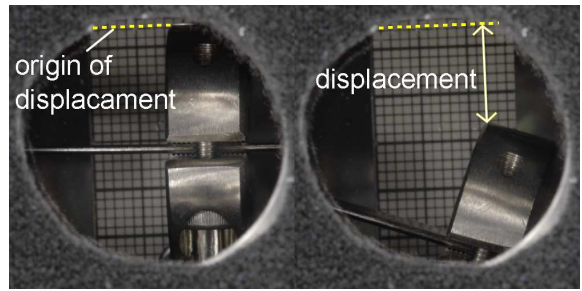}
  \caption{The way to measure displacement: left one is a photo taken before hydrogen was introduced.  }
  \label{displacement}
 \end{center}
\end{figure}

\begin{figure}[!tb]
 \begin{center}
  \includegraphics[width=6cm,clip]{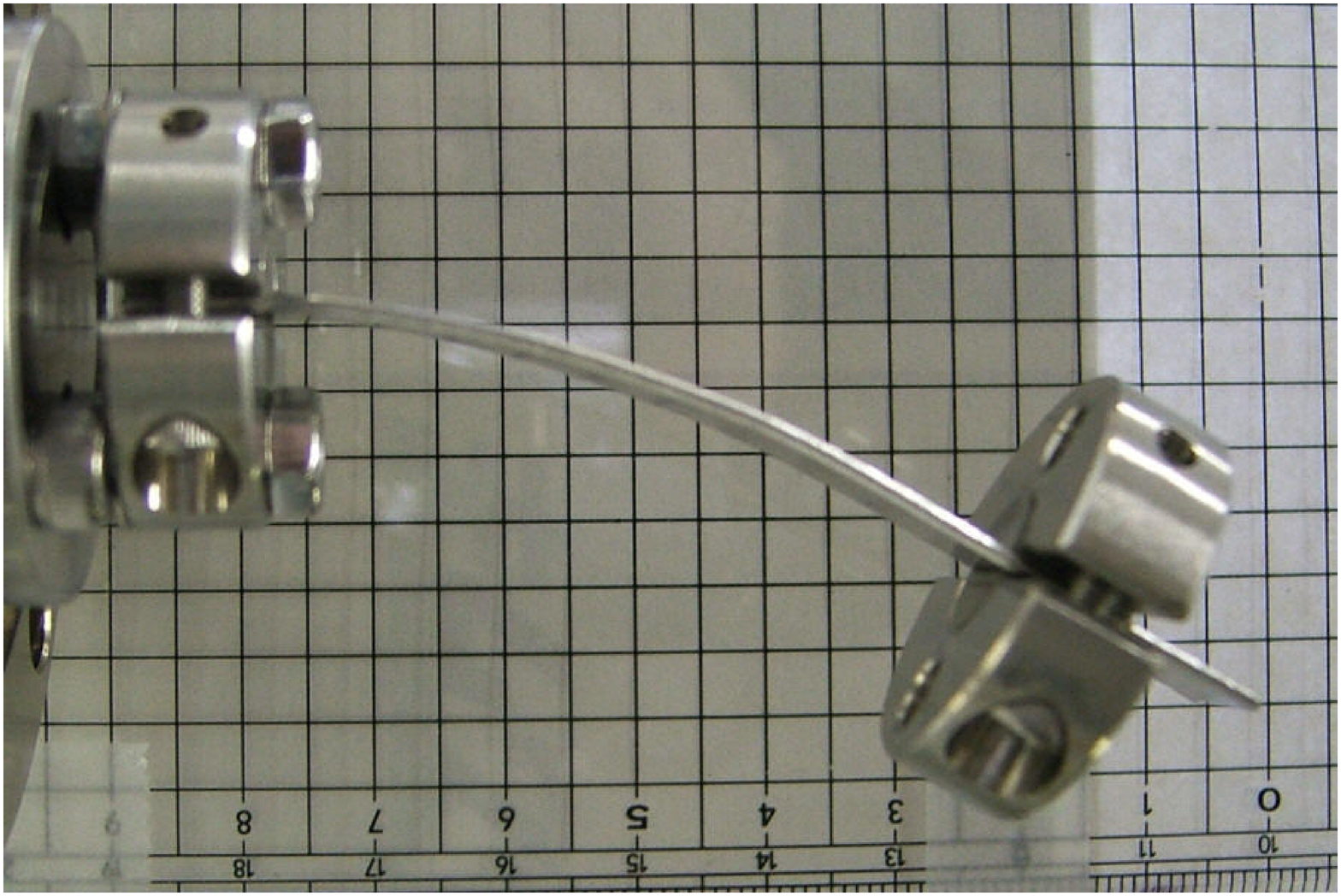}
  \caption{The Pd plate used for Experiment B1:  The scaling pad behind the plate had a 5 mm mesh.}
  \label{plate0318}
 \end{center}
\end{figure}

The results of two experiments, which we call B1 and B2, are shown as follows.  Experiment B1 was aimed to see the drastic change, whereas Experiment B2 was for observing detailed time dependence of the bending in relatively slow absorption rate.  The pressure inside the vessel, the displacement of the plate and photos of the plate at some time points for Experiment B1 and B2 are shown in Figs.~\ref{Pressure0218} and \ref{Pressure0318}, respectively.  In each figure, the upper half shows the displacement and the lower half describes the pressure in the vessel.  The displacement, which was recorded all the time throughout an experiment, was determined by measuring the position of a corner of the jig as shown in Fig.~\ref{displacement}. The positive direction is defined as vertically upward.  By assuming the bend of the plate was uniform, the relative strain of the plate $\epsilon$, which is defined as $\epsilon = ({\rm displacement})/({\rm original\ length})$, was calculated as the relative elongation of the upper surface of the plate when the 6 cm distant point of the center plane from the fixed moves by the measured displacement with uniform bending without elongation.  However, as shown in Fig.~\ref{plate0318}, some part of a plate was actually bent more than other part after the procedure.  The blue and green dashed lines show the time when the introduction and the exhaust of hydrogen started.  The pressure after the evacuation started is drawn by a red curve, whose scale is given on the right side.  

Large displacements were observed both at the absorption and desorption processes, but they occurred mostly in short limited times. The most surprising was result B1.  Especially large was the displacement during the first two minutes: 5.2 mm ($\epsilon=1.8 \times 10^{-3}$), which was the 30\% of the total distortion in a cycle.  No change was observed 20 minutes after the hydrogen was introduced or later.  After the desorption of hydrogen started, it started being bent again.  However, the speed of displacement was smaller in the desorption stage, presumably because the speed of the desorption was slower than that of the absorption.  

The motion of the plate in Result B2 was different from that for B1.  The displacement during the first two hours was relatively small, and the main distortion occurred during the next two hours, which means that the speed of the change in shape increased.  At the desorption phase, the plate went upward, whereas it continued to go down in Experiment B1.  

The shearing force applied to the plate by the weight was 1.22N.  Ordinary bending by this amount of stress calculated from a beam theory formula $\delta_h = 4FL^3/Ebh^3$, where F, L, b and h are force, length, width and height of a beam, with the Young's modulus of palladium, $1.10 \times 10^{11}$N/m$^2$ at room temperature, is 0.9 mm. (This estimation is valid for the 150$^{\circ}$C case, because the Young's modulus changes by less than 2\% due to this temperature difference.\cite{Young}) In addition, raising the temperature to 150 $^{\circ}$C induced at most a 1 mm displacement to the plate, and there was only one case for this kind of distortion among several experiments.  The displacement by a cycle of absorption and desorption of hydrogen was from 9.4 mm to 17.1 mm.   Therefore, hydrogen caused a deformation that was 10-times larger than the elastic bend.

\subsection{Discussion: Bend}
\begin{figure}[!tb]
 \begin{center}
  \includegraphics[width=7.5cm,clip]{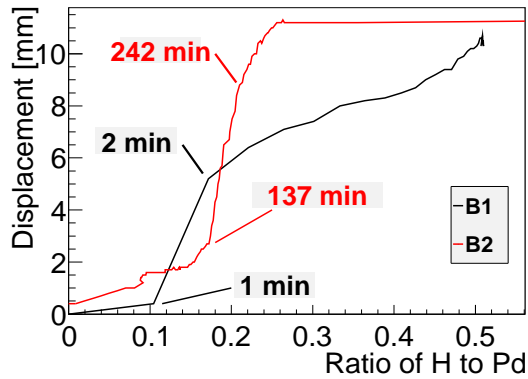}
  \caption{Relation between ${\rm x=H/Pd}$ and Displacement: The time written in the graph is the time that passed since the hydrogen was introduced to the vessel.  }
  \label{xd}
 \end{center}
\end{figure}

The most remarkable finding from these experiments is that the bending deformation occurred rather suddenly, in a few minutes in Experiment B1, at a certain point of hydrogen absorption and desorption. We studied how this transition timing is related to the hydrogen concentration.  Fig.\ref{xd} shows the relation between the portion of hydrogen atom and the deformation.  This clearly shows that the deformation occurs fastest when the concentration of hydrogen atom in palladium metal is from 10\% to 20\%.  This ratio corresponds to $\alpha + \beta$ phase.  Since the displacement increases as hydrogen increases in palladium, the region of the $\beta$ phase gets larger during this change.  This implies the deformation is related to superplasticity in the naive sense, which ocurrs when two phases coexists and one is changing to the other.  

When the deformation was fastest in Experiment B1, its rate of the elongation is $2.8 \times 10^{-5}$s$^{-1}$.  This is comparable to the various flow rate and stress appearing in Ref.\cite{GJ1965}, but the direction of stress in our experiment was different from other experiments; in ordinary superplasticity experiments one applies tension to a metal, but the plate experienced shearing stress in our experiment.  This should be one of the reasons we observed much larger elongation rate than Beeri et al.'s result\cite{Beeri2009}, $5\times 10^{-7}$s$^{-1}$.  Another factor is the hydrogen absorption rate due to the difference in the temperature, 20 minutes in Experiment B1 and 10 hours in Ref.\cite{Beeri2009}, but the stress in the different direction has a large effect.  The magnitude of the stress was smaller by a factor of from 100 to 1000 than the stress applied in Ref.\cite{Beeri2009}.  The importance of the direction of the stress is also true to Experiment B2, for it has fastest elongation rate of $8.7\times 10^{-7}$s$^{-1}$ that is comparable to that in Ref.\cite{Beeri2009} with much smaller stress than Ref.\cite{Beeri2009}.  This implies that the deformation is caused by a shearing stress with a magnitude much smaller than tension case.  This bending speed and bending larger than the elastic deformation might imply the superplasticity in a naive sense, but since total strain is not as large as 100\%, there is not any evidence strong enough to support superplasticity.

\section{Dynamical Measurements of the Extension of Pd Plates} 

\subsection{Experimental Procedure: Extension}

\begin{figure}[!tb]
 \begin{center}
  \includegraphics[width=7.5cm,clip]{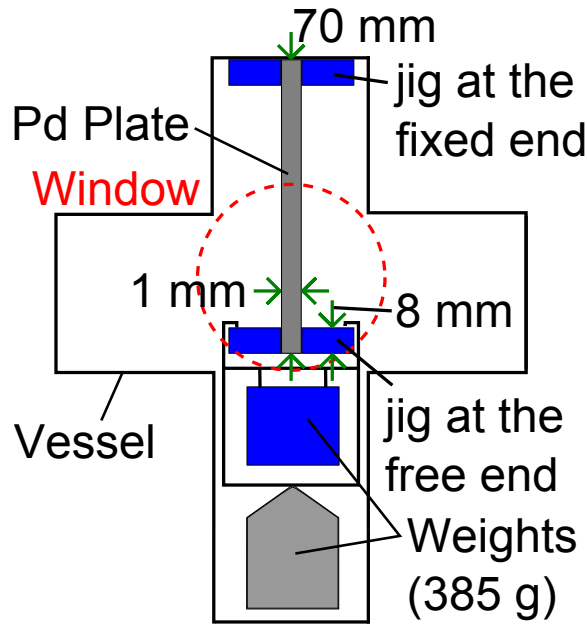}
  \caption{The setting for the experiments to measure the extension }
  \label{Vertical}
 \end{center}
\end{figure}

The setting of the experiments to measure the extension of the palladium plate is shown in Fig.~\ref{Vertical}.

A plate of palladium was hung vertically in a vacuum vessel with its largest surface parallel to the line of sight.  The size of the plate was the same as in the previous experiments: $1\ {\rm mm}  \times 10\ {\rm mm} \times 70\ {\rm mm}$.  A piece of kapton tape was attached to mark the point 6 cm distant from the fixed edge.  The scaling pad same as the one used in the bending experiments was located 7.5 mm behind the plate.  A 385 g weight, which caused 377 kPa tension, was hung below the plate.  The configuration of the camera and the way to heat the vessel were the same as in the experiments to measure the bend.  

The procedure of one cycle of absorption and desorption was the same as the experiments to measure the bend.  Since a larger vessel was used in these experiments in order to hang heavier weights, it took longer for the vessel to reach to 150 $^{\circ}$C.  

\subsection{Experimental Results:  Extension}
\begin{figure}[!tb]
 \begin{center}
  \includegraphics[width=8.5cm,clip]{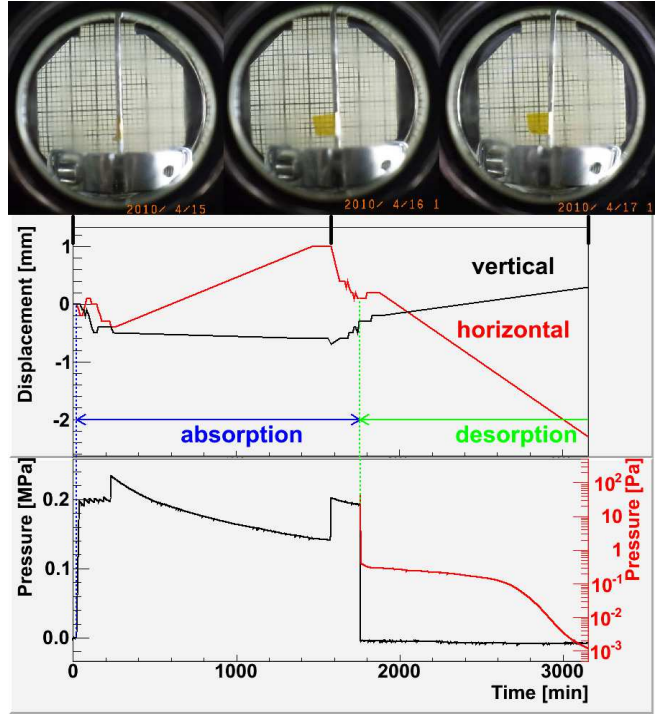}
  \caption{Result E1; with weight, 0.2 MPa H$_2$}
  \label{Pressure0413}
 \end{center}
\end{figure}

\begin{figure}[!tb]
 \begin{center}
  \includegraphics[width=8.5cm,clip]{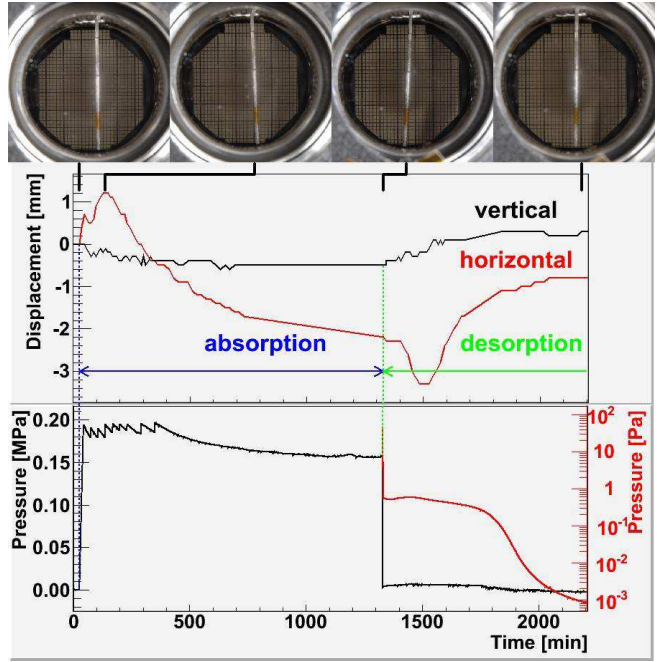}
  \caption{Result E2; without weight, 0.2 MPa H$_2$}
  \label{Pressure0320}
 \end{center}
\end{figure}

The results of two experiments, with and without the weight, which are called E1 and E2, respectively, are shown in Figs.~\ref{Pressure0413} and \ref{Pressure0320}.  The plates used at both experiments were annealed before the experiments, but before annealing they had experienced a couple of cycles of absorption and desorption of hydrogen.  The pressure was depicted in the same way as in the experiment to measure the bending.  In the upper half, the black line shows the vertical displacement and the red line indicates the horizontal one.  The displacements were measured as the change in the position of the point on the plate 6 cm distant from the fixed edge.  The positive value in the horizontal displacement is defined as for the right direction and that in the vertical displacement is for the upward direction, thus corresponding to a negative strain $\epsilon$ (shrinking).  The inner volume of the vessel that was filled with gas was 494 cm$^3$ during Experiment E1 and 512 cm$^3$ during Experiment E2.  

The extension of the plate after one cycle of absorption and desorption was not very drastic compared to the bend.  The observed strain was at most 0.5mm.  This is much larger than the elongation calculated by a formula $\delta _v =FL/Ebh$ from Young's modulus, $2 \times 10^{-4}$ mm for a 6 cm long plate.  The relative strain $\epsilon$ was therefore $8 \times 10^{-3}$ or smaller.  
However, after the desorption, both at Result E1 and at result E2, the metal shrank by 0.3 mm ($\epsilon = -5 \times 10^{-3}$).  This implies that the 377 kPa weight did not have a significant effect on the elongation of the plate, as the extension during the absorption can be explained by the lattice expansion.  

Results E1 and E2 show that the plate become warped during the absorption and desorption of hydrogen.  In both cases, the plate was bent both to the left and to the right.  This made the strain estimated above vague, because the curved plate have different elongation in two sides and because both horizontal and vertical displacement requires more elongation than the simple vertical strain.  The maximum horizontal displacement was larger during the experiment without weight.  

\subsection{Discussion: Extension}
The results of experiments E was consistent with the previous report\cite{Beeri2009}.  Smaller stress applied to the metal in this experiment explains the following feature; the strain of the plate during the absorption was smaller and shrink at the end of the desorption was larger than that in Ref.\cite{Beeri2009}.  In addition, qualitative behavior also matches another report\cite{JpnAcad}; the plate shrank after one cycle of the absorption and the desorption.  

The most remarkable feature of the present experiment is the anomalous warping behavior of the plate.  If we regard the warping as deformation that breaks the shape symmetry, examples of such spontaneous deformation of Pd metals is reported in Ref.\cite{JpnAcad}; Pd metals of various shapes change their shape in the direction of minimizing the surface area.  The observed warp itself is consistent with this tendency, as far as the surface area was decreasing by the warping.  However, we do not understand yet why the direction of the warping changes in the middle of absorption or desorption.  Faster absorption from one surface seems to explain the warping, as higher hydrogen density results in the expansion of the lattice.  However, since the horizontal displacement changed its sign during the absorption in both E1 and E2, which means that the surface with slower absorption rate had larger final expansion, this simple interpretation cannot explain the whole behavior.  

The larger warp in the Experiment E2, which was without the weight, suggests that the weight prevented the plate from curving.  Combining these with the result of Experiment B, we came to conclusion that palladium plate was easily affected by shearing force than by tension.  It is true that this is the behavior same as that of the elastic deformation of a beam.  However, when we think whether the weight induced a deformation significantly larger than elastic deformation, this result is not trivially derived from beam theory.  

\section{Conclusion}
We observed a spectacular displacement of a palladium plate caused by a small external shearing stress during absorption and desorption of hydrogen at 150$^{\circ}$C. This took place in a short time during absorption, and also during desorption. This rapid change, which happened around $x \approx 0.1 - 0.2$ where x is the ratio of H to Pd atom, seemed correlated with a phase transition of PdH$_{\rm x} $from $\alpha$ phase to $\alpha + \beta$.  In contrast to the large deformation when the plate was held horizontally, there was nearly no effect when it was hung vertically.  This is consistent with previous reports, but we observed warping motion of the plate, which remains to be explained.  Comparison of two kinds of experiments showed that the palladium plate changes its shape more easily by shearing stress than by tension.  In order to comprehend this phenomenon further, various microscopic analyses in real time are necessary, such as x-ray diffraction, NMR, $\mu$SR and neutron diffraction.  This would reveal if the observed deformation is because of superplasticity.

\section{Acknowledgements}
This research is supported by the Strategic Program for R\&D of RIKEN.  We would like to thank Dr. K. Itahashi for providing us with photographing equipment, Dr. M. Sato for writing data acquisition and Dr. S. Kamiguchi and Chemical Analysis Team at RIKEN for allowing us to use instruments for annealing.  We are grateful to Dr. Nishimura with insightful discussion and Prof. M. Iwasaki and Prof. R. S. Hayano for the stimulating support.

\bibliographystyle{model1a-num-names}

\begin{thebibliography}{99}
\bibitem{Pd-H} F. A. Lewis, The Palladium Hydrogen System, Academic Press, 1967 
\bibitem{Phase} A. G. Knapton, Plat. Met. Rev. {\bf 21} (1977) 44-50
\bibitem{HinM}E. Wicke and H. Brodowsky, in: G. Alefeld and J. Volkl (Eds.) Hydrogen in Metals II, Springer, 1978, pp.81
\bibitem{alpha} K. Skold and G. Nelin, J. Phys. Chem. Solids {\bf 28} (1967) 2369-2380
\bibitem{beta}J. E. Worsham, M. K. Wilkinson, and C. G. Shull, Phys. Chem. Solids {\bf 3} (1957) 303-310
\bibitem{Sugeno} T. Sugeno and H. Kawabe, Mem. Inst. Scient. Ind. Res. Osaka Univ. {\bf 14} (1957) 25-35
\bibitem{OldPaper} W. Krause and L. Kahlenberg, Trans. Electrochem. Soc. {\bf 68} (1935) 449
\bibitem{PMR} J. B. Hunter, Platin. Metals Rev. {\bf} 4 (1960) 130-131
\bibitem{ST}  A. Kufudakis and J. Cermak, Surf. Technol. {\bf 16} (1982) 57-66
\bibitem{IJHE}  R. V. Kotelva and J. L. Glukhova, Int. J. Hydrogen Energy {\bf 22} (1997) 175-177
\bibitem{AM} Z. R. Xu and R. B. Mclellan, Acta Mater. {\bf 46} (1998) 4543-4547
\bibitem{JSMS1} Y. Jung, H. Suehiro and Y. Sakai, J. Soc. Mat. Sci. Japan {\bf 49} (2000) 1242-1248 (in Japanese)
\bibitem{JSMS2} Y. Jung and Y. Sakai, J. Soc. Mat. Sci. Japan {\bf 50} (2001) 999-1006 (in Japanese)
\bibitem{JpnAcad}T. Yamazaki, M. Sato, S. Itoh, Proc. Jpn. Acad., Ser B {\bf 85} (2009) 183-186
\bibitem{Young}H. Masumoto, H. Saito and S. Kadowaki, Sci. Rep. of the Res. Inst., Tohoku Univ. Ser. A, Phys., Chem. and Meall., {\bf 19} (1967) 294-303
\bibitem{Gortsov1981} V. A. Gortsov, Mat. Sci. Eng. {\bf 49} (1981) 109-125
\bibitem{Beeri2009}O. Beeri and D. C. Dunand, Mat. Sci. Eng. A {\bf 523} (2009) 178-183
\bibitem{Morreale2003}B. D. Morreale, M. V. Ciocco, R. M. Enick, B. I. Morsi,
B. H. Howard, A. V. Cugini, K. S. Rothenberger, J. Membrane Sci. {\bf 212} (2003) 87-97
\bibitem{GJ1965}G. W. Greenwood and R. H. Johnson, Proc. R. Soc. Lond. A {\bf 283} (1965) 403-422
\end{thebibliography}

\end{document}